\documentclass{kluwer}    
\usepackage{graphicx}

\newdisplay{guess}{Conjecture}

\begin{document}                                                                                   
\begin{article}
\begin{opening}         
\title{Solar-like oscillations in semiregular variables}

\author{Timothy R. \surname{Bedding}}
\runningauthor{T.R. Bedding}
\runningtitle{Solar-like oscillations in red giants}
\institute{School of Physics, University of Sydney 2006, Australia}

\begin{abstract}
Power spectra of the light curves of semiregular variables, based on visual
magnitude estimates spanning many decades, show clear evidence for
stochastic excitation.  This supports the suggestion by
Christensen-Dalsgaard et al.\ (2001) that oscillations in these stars are
solar-like, i.e., stochastically excited by convection, with mode lifetimes
ranging from years to decades.
\end{abstract}

\end{opening}           

Oscillating red giants with high luminosity -- the long period variables --
are conventionally divided into Miras and semiregulars.  Both can be
monitored visually, thanks to the extreme temperature sensitivity of the
TiO absorption bands that dominate the visible spectrum.  For some stars,
visual magnitude estimates by amateur astronomers span many decades.

Mira variables have large amplitudes and are very regular, reflecting the
nature of the driving process, which is self-excitation via opacity
variations.  Semiregulars (SRs), on the other hand, have lower amplitudes,
less regularity and often show two or three periods \cite{KSC99,WAA99}.  In
these stars, it seems plausible that there is a substantial contribution
from convection to the excitation and damping.  Indeed,
\citeauthor{ChDKM2001} (\citeyear{ChDKM2001}) have suggested that the
amplitude variability seen in SRs is consistent with the pulsations being
solar-like, i.e., stochastically excited by convection.  The subject of
Mira-like versus solar-like excitation has also been discussed in the
context of K~giants by \citeauthor{DGH2001} (\citeyear{DGH2001}).

\begin{figure}

\centerline{\hspace*{6mm}\scalebox{0.6}[0.6]{\includegraphics{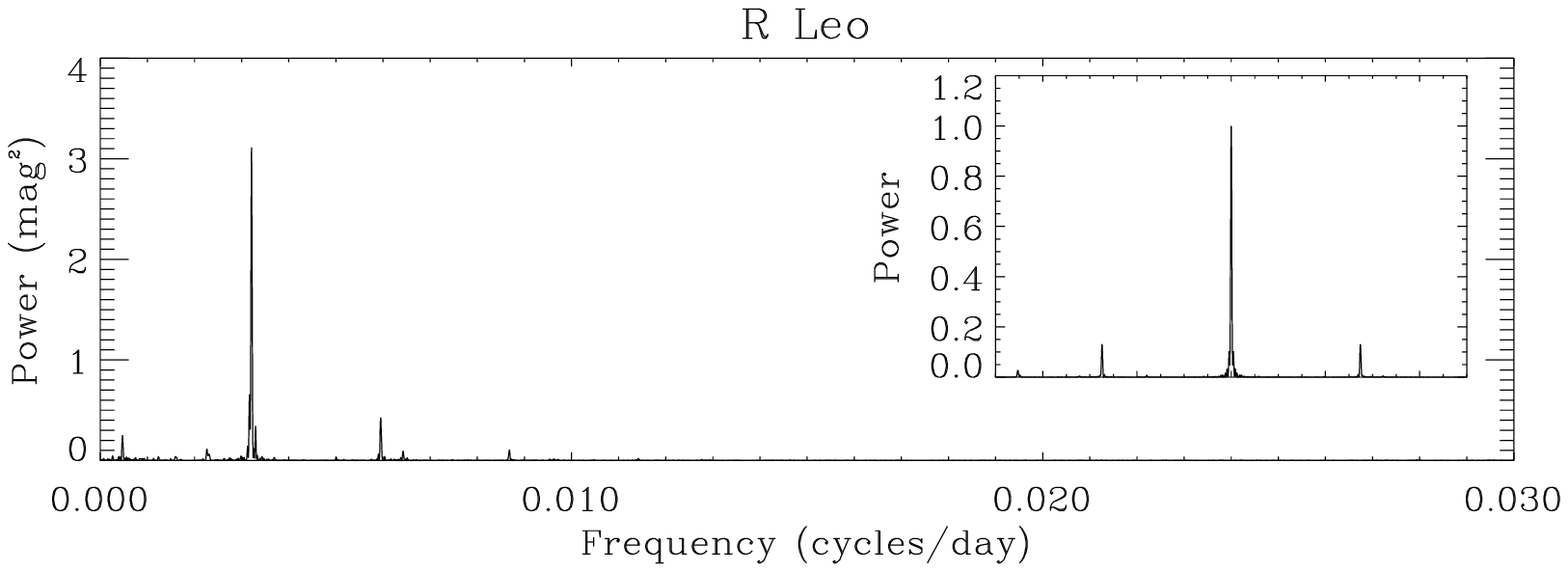}}}
\medskip

\centerline{\hspace*{1.8mm}\scalebox{0.6}[0.6]{\includegraphics{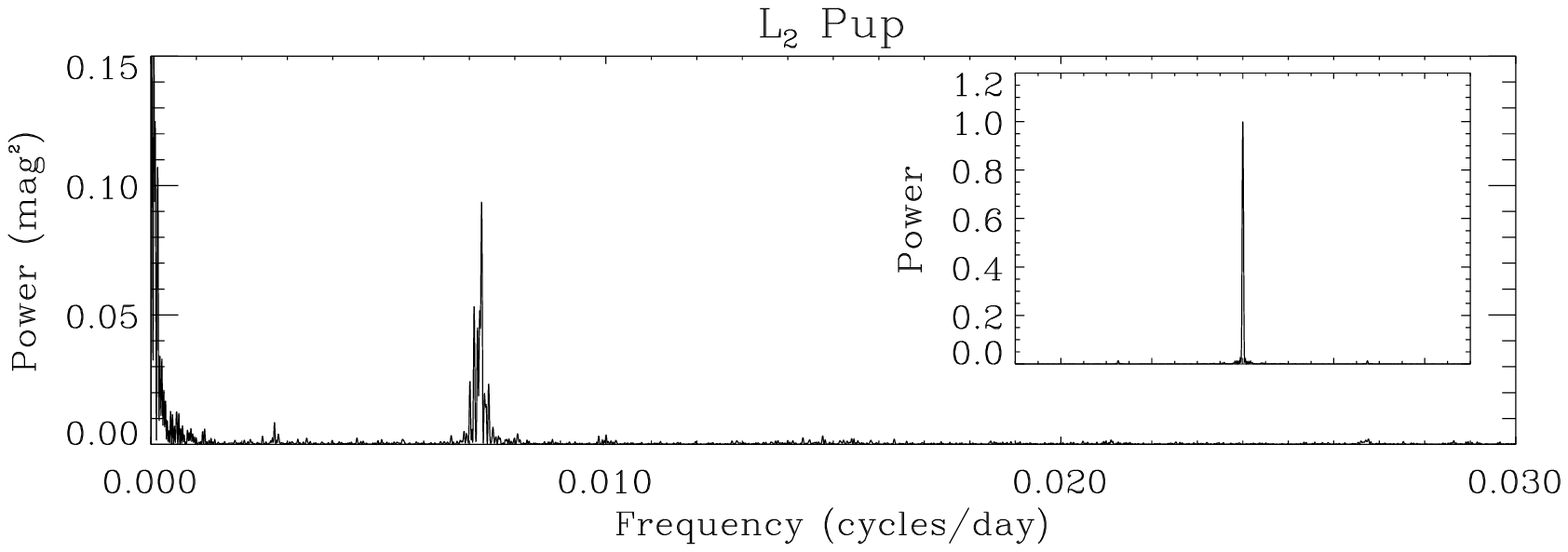}}}
\medskip

\centerline{\scalebox{0.6}[0.6]{\includegraphics{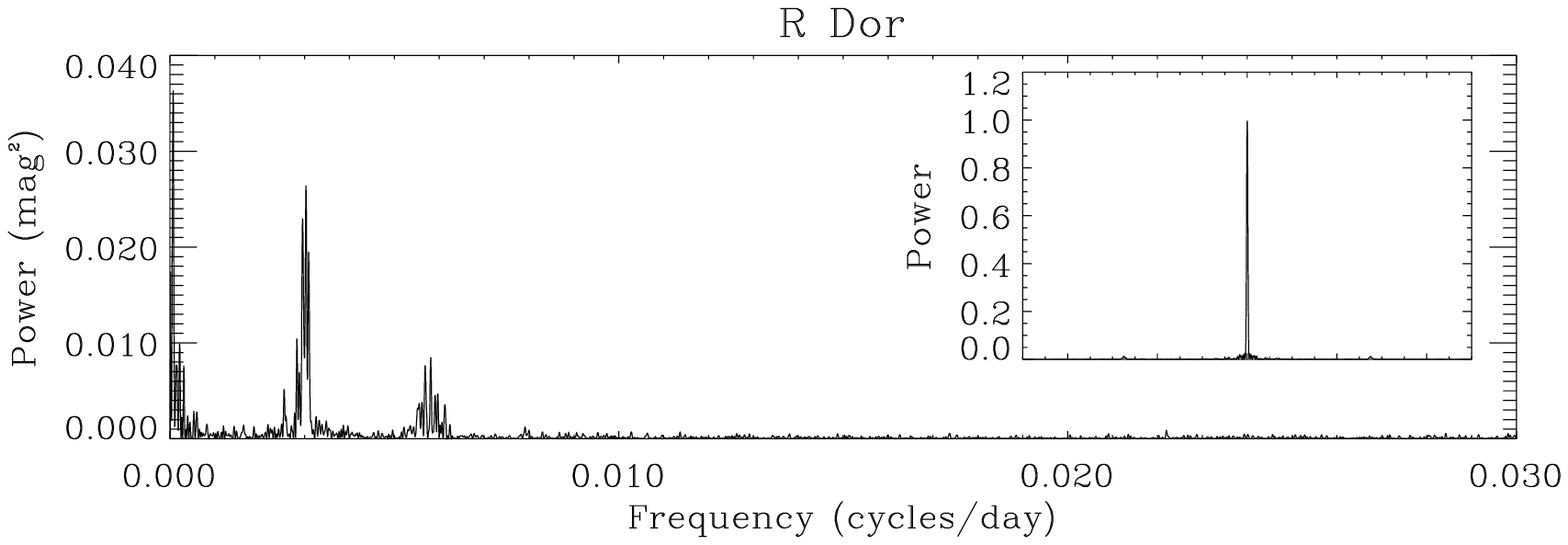}}}
\medskip

\centerline{\scalebox{0.6}[0.6]{\includegraphics{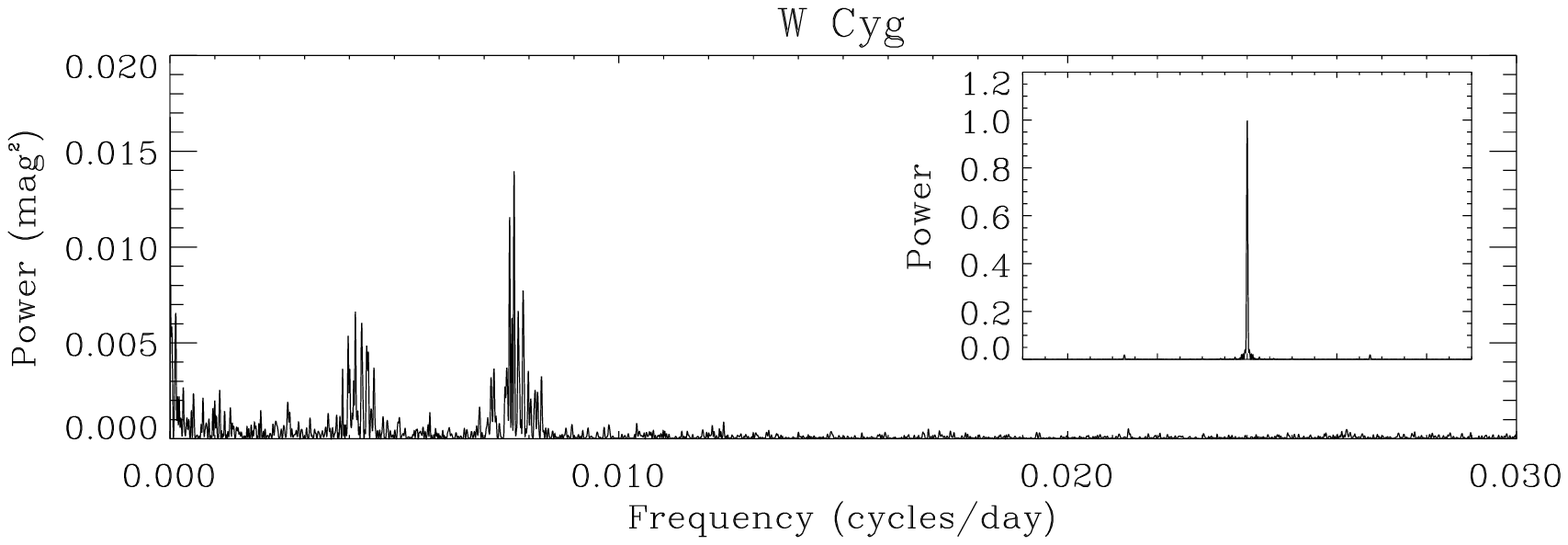}}}
\medskip

\centerline{\scalebox{0.6}[0.6]{\includegraphics{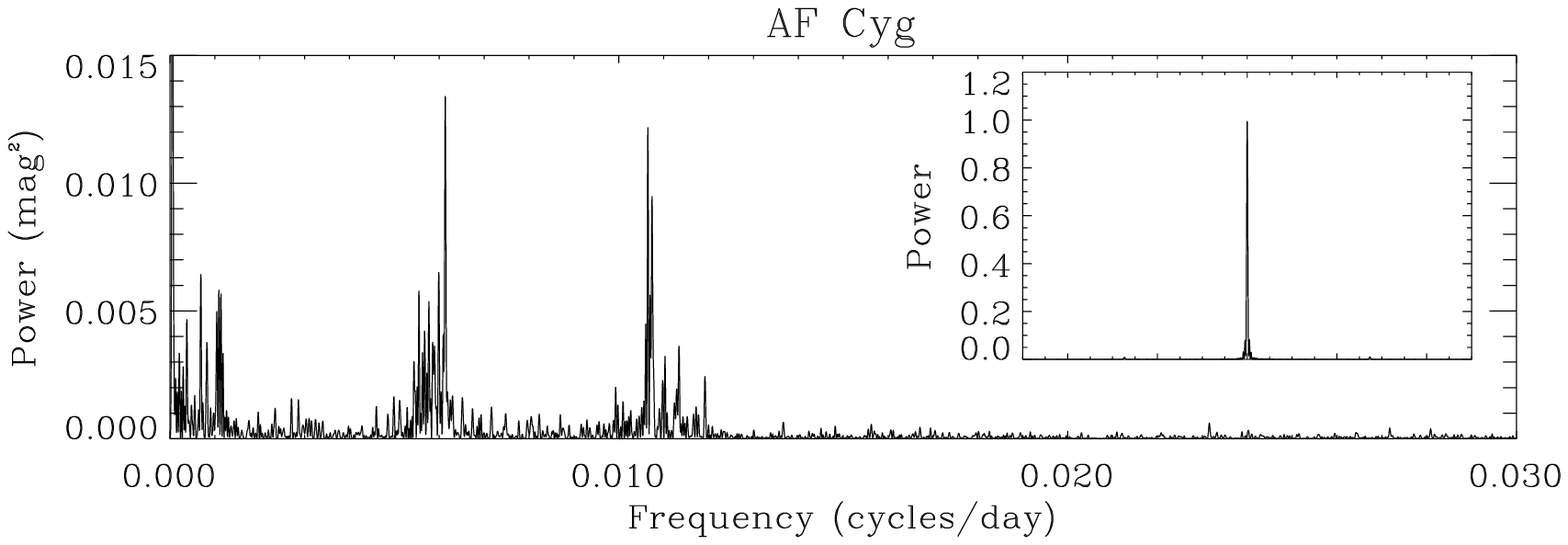}}}

\caption[]{\label{fig.power} Power spectra of visual observations of a Mira
and four semiregulars.  In each case, the inset shows the spectral window.}
\end{figure}

\begin{figure}

\centerline{\scalebox{0.6}[0.6]{\includegraphics{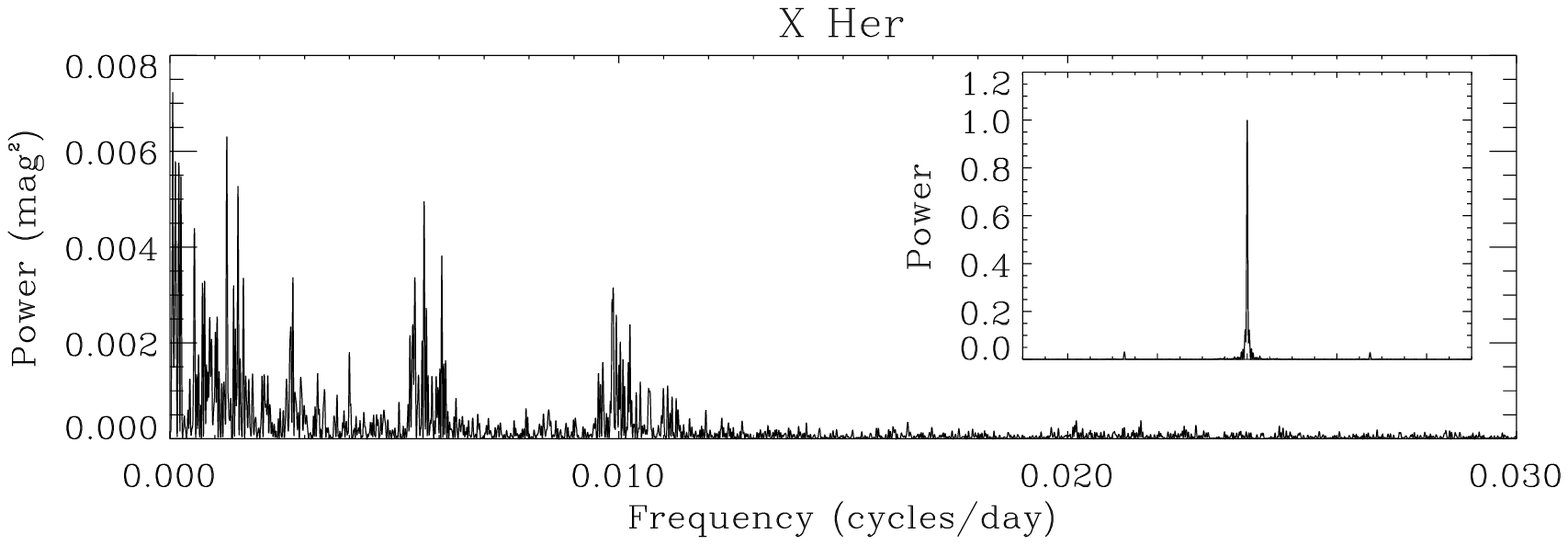}}}
\medskip

\centerline{\scalebox{0.6}[0.6]{\includegraphics{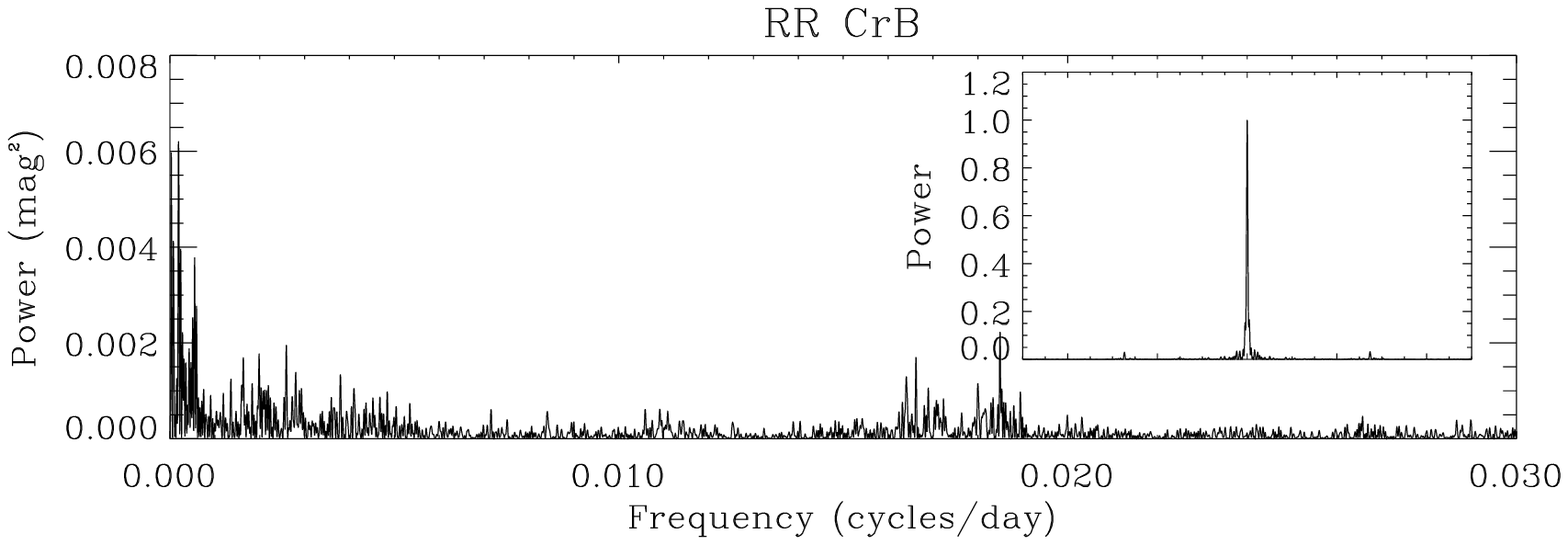}}}
\medskip

\centerline{\scalebox{0.6}[0.6]{\includegraphics{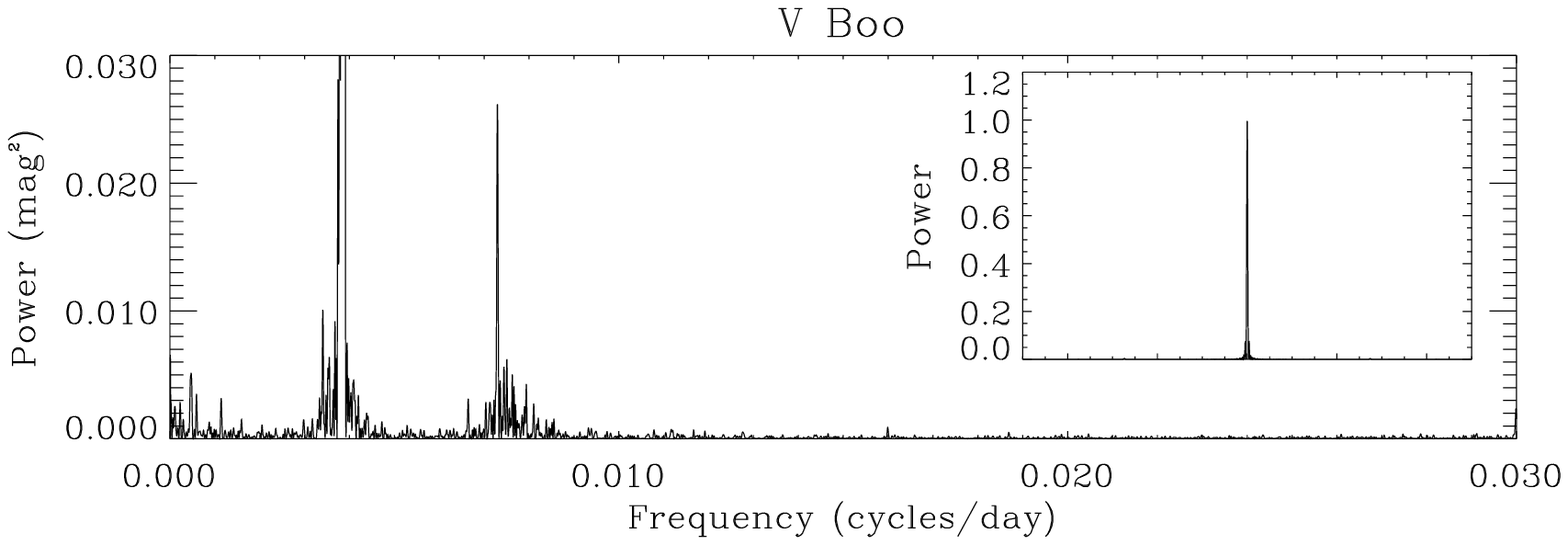}}}
\medskip

\centerline{\scalebox{0.6}[0.6]{\includegraphics{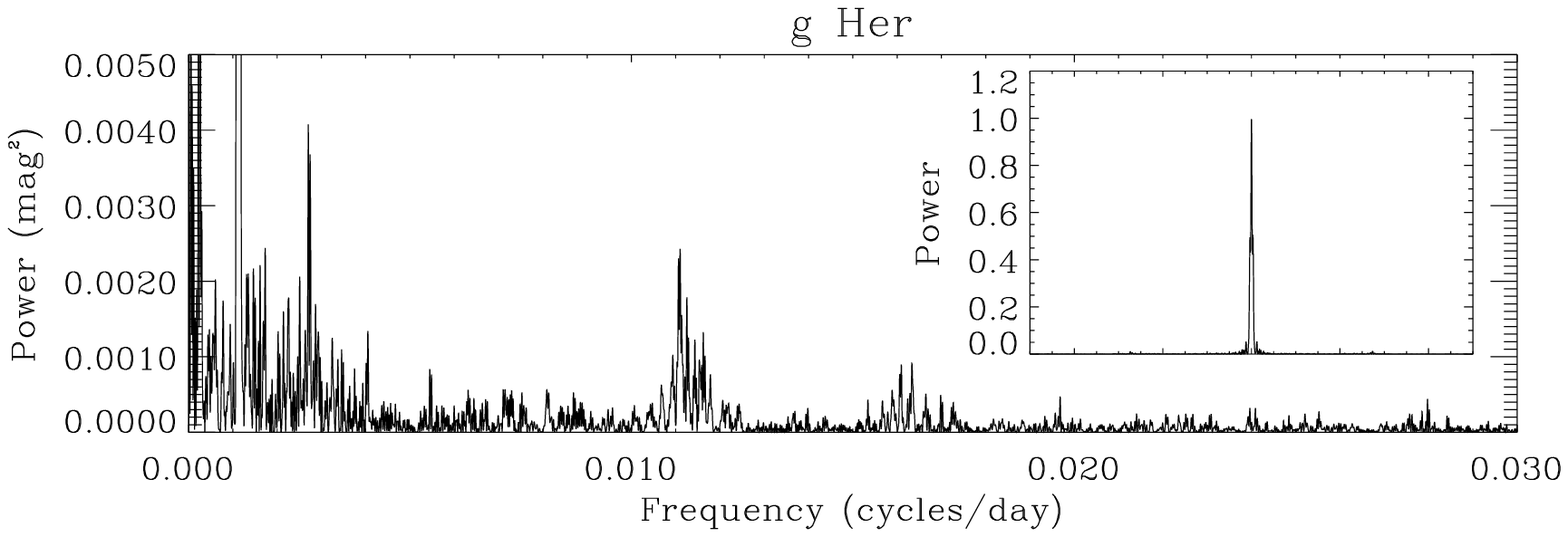}}}
\medskip

\centerline{\includegraphics[width=11.2cm]{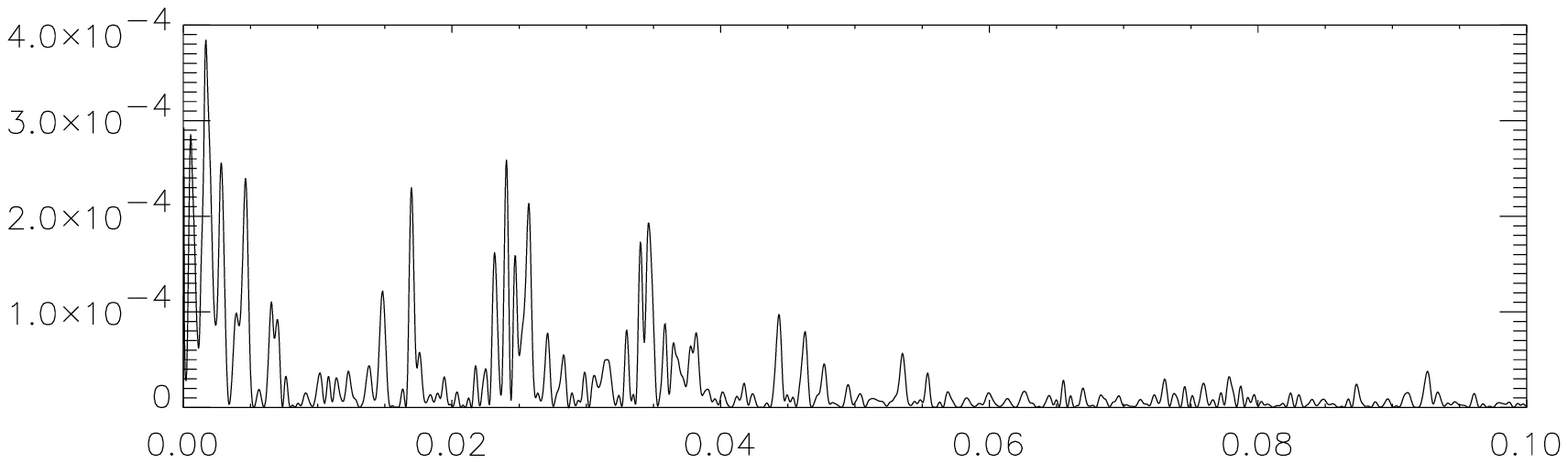}}

\caption[]{Same as Fig.~1, for four more semiregulars, plus one red giant
from MACHO observations of the LMC (bottom panel; note change of horizontal
scale).}
\end{figure}

Figures~1 and~2 show power spectra of visual observations for some of the
best-studied long period variables.  The first star, R~Leo, is a typical
Mira (period 310\,d) and shows a narrow peak in the power spectrum.  This
indicates that the pulsation is stable in both period and phase.

The other stars are SRs, and all show strong evidence for
stochastic excitation.  The power from each oscillation mode is split into
a series of peaks under a narrow envelope.  This structure is typical of a
stochastically excited oscillator and is strikingly similar to close-up
views of individual peaks in the power spectrum of the Sun \cite{T+F92}.
We can estimate the mode lifetime from the width of the envelope.  This
seems to range from a few decades (L$_2$~Pup) down to only a few years
(e.g., X~Her).  For most doubly-periodic stars, the mode lifetime is
similar for both periods.

RR~CrB appears to show two closely space modes (periods 54 and 60\,d),
which at first sight seem hard to understand as consecutive low-order
radial modes.  However, observations of semiregulars both locally
\cite{KSC99} and in the LMC \cite{WAA99} also show some stars with period
ratios close to 1.1, and models by \inlinecite{WAA99} indicate plausible
identifications with low-order modes.

V~Boo is an unusual star, with a Mira-like mode (258\,d) whose amplitude
has decreased steadily over the past 90 years, plus a shorter-period mode
(137\,d) that has remained relatively constant in amplitude
\cite{SGK96,BZJ98,KSC99}.  As expected, the power spectrum at the long
period is very strong (the peak is way off scale, at 0.1\,mag$^2$), while
the spectrum around the short period shows a low broad hump.  However,
there is also narrow peak just left of centre in the latter.  This peak
does not coincide with the harmonic of the longer period, and apparently
indicates a coherent long-lived component to an otherwise stochastically
excited oscillation.

g~Her (= 30~Her) has two pulsation modes with solar-like envelopes (90 and
60\,d), but also a much longer period (890\,d) that is coherent (the peak
is off the top of the graph, at 0.033\,mag$^2$).  The latter is a typical
example of a long secondary period, often seen in SRs and probably due to
binarity (\opencite{WAA99}; Huber et al., these Proceedings).

Finally, the bottom panel of Fig.~2 shows the power spectrum of a red giant
in the LMC, based on seven years of data from the MACHO database.  We see
evidence for perhaps as many as five equally-spaced modes, all with similar
envelopes.  There is clearly much to be learned from data such as these.

\end{article}

\begin{thebibliography}{}

\bibitem[\protect\citeauthoryear{Bedding \& Zijlstra}{1998}]{B+Z98}
Bedding, T.~R., \& Zijlstra, A.~A., 1998, ApJ, 506, L47.

\bibitem[\protect\citeauthoryear{Bedding et~al.}{1998}]{BZJ98}
Bedding, T.~R., Zijlstra, A.~A., Jones, A., \& Foster, G., 1998, MNRAS, 301,
  1073.

\bibitem[\protect\citeauthoryear{Bedding et al.}{2002}]{BZR2002}
Bedding, T.~R., Zijlstra, A.~A., Retter, A., Yamamura, I., Jones, A.,
  Whitelock, P., Marang, F., \& Matsuura, M., 2002, MNRAS (in press).

\bibitem[\protect\citeauthoryear{Christensen-Dalsgaard et
al.}{2001}]{ChDKM2001}
Christensen-Dalsgaard, J., Kjeldsen, H., \& Mattei, J.~A., 2001, ApJ, 562,
L141.

\bibitem[\protect\citeauthoryear{Dziembowski et al.}{2001}]{DGH2001}
Dziembowski, W.~A., {Gough}, D.~O., {Houdek}, G., \& {Sienkiewicz}, R., 2001,
  MNRAS, 328, 601.

\bibitem[\protect\citeauthoryear{Kiss et~al.}{1999}]{KSC99}
Kiss, L.~L., {Szatm{\' a}ry}, K., {Cadmus}, R.~R., \& {Mattei}, J.~A., 1999,
  A\&A, 346, 542.

\bibitem[\protect\citeauthoryear{Szatm\'ary et al.}{1996}]{SGK96}
Szatm\'ary, K., G\'al, J., \& Kiss, L.~L., 1996, A\&A, 308, 791

\bibitem[\protect\citeauthoryear{Toutain \& {Fr\"ohlich}}{1992}]{T+F92}
Toutain, T., \& {Fr\"ohlich}, C., 1992, A\&A, 257, 287.

\bibitem[\protect\citeauthoryear{Wood et al.}{1999}]{WAA99} 
Wood, P.~R., {Alcock}, C., {Allsman}, R.~A., et al., 1999, in IAU Symp.\
191: Asymptotic Giant Branch Stars, eds.\ T. Le Bertre, A. Lebre \& C.
Waelkens, p.~151
  
\end{thebibliography}
\end{document}